\documentclass[onecolumn,authoryear]{els-mrw} 

\usepackage{amsmath,amssymb,amsfonts,amsthm,makeidx,graphicx}
\usepackage{aas_macros}
\usepackage{txfonts}
\usepackage{helvet}
\usepackage{hyperref}

\begin{document}

\chapter{Population Synthesis of Gravitational Wave Sources}\label{chap5}

\author[1]{Katelyn Breivik}%


\address[1]{\orgname{Carnegie Mellon University}, \orgdiv{McWilliams Center for
Cosmology and Astrophysics, Department of Physics}, \orgaddress{Pittsburgh, PA
15213, USA}}

\articletag{Chapter Article tagline: update of previous edition,, reprint..}

\maketitle


\begin{glossary}[Nomenclature] \begin{tabular}{@{}lp{34pc}@{}} 

  AGN &active galactic nucleus\\
  BH &black hole\\
  BSE &binary stellar evolution\\
  EM &electromagnetic\\
  DCO &double compact object\\
  GC &globular cluster\\
  GW &gravitational wave\\
  LIGO &Laser Interferometer Gravitational Wave Observatory\\
  LISA &Laser Interferometer Space Antenna\\
  NS &neutron star\\
  NSC &nuclear star cluster\\
  SSE &single stellar evolution\\
  WUMa &W Ursula Majoris\\
  WD &white dwarf\\

\end{tabular} \end{glossary}

\begin{abstract}[Abstract]
The simulation of gravitational wave source populations and their progenitors is
an endeavor more than eighty years in the making. This is in part due to a wide
variety of theoretical uncertainties that must be taken into account when
describing how stellar populations evolve over cosmic time to produce double
stellar remnant binaries. Population synthesis software has been developed as a
means to investigate these uncertainties under a wide variety of physical
assumptions and stellar population formation environments. In this chapter we
discuss the development history of population synthesis software with a special
focus on work aimed at understanding the formation of gravitational wave
populations. We detail the assortment of population synthesis tools in use today
that simulate GW populations which are born and evolve in different
astrophysical environments. We further discuss the GW population rates and
features associated with each environment that have been predicted for both
ground and space-based GW detectors. We finish with considerations of future
work that combines possible constraints from electromagnetic surveys that may
provide key findings that break current degeneracies in population synthesis
predictions of GW source populations. 
\end{abstract}

\begin{BoxTypeA}[]{Key points}
  \begin{itemize}
  \item A central use of population synthesis algorithms and codes prior to the
  detection of gravitational waves was the prediction of potential stellar
  remnant populations that could be detected by, and thus serve as funding
  motivation for, both ground and space-based gravitational wave detectors. 
  \item Gravitational wave sources can originate in a variety of astrophysical
  environments which are delineated between isolated binary systems, triple star
  systems, stellar clusters of varying size and density, and gaseous disks which
  could occupy the centers of galaxies or be created by non-accreted binary
  material.
  \item The rates of gravitational wave sources vary widely both due to the
  source type and environments, but nearly all are overlapping. Features in the
  observed parameters of gravitational wave source populations in different
  environments may break the rate degeneracies. 
  \item Combining electromagnetic observations of binary-star populations with
  one stellar remnant and one fusing star with gravitational wave observations
  may also break degeneracies in the predictions for gravitational wave
  populations from different astrophysical environments.
  \end{itemize}
  \end{BoxTypeA}

\section{Introduction}\label{chap5:sec1} Stellar populations and the remnants
they produce have been known as potential gravitational wave (GW) sources since
the 1960's. Calculations for the emission of GWs \citep{1963PhRv..131..435P},
their effect on the orbital evolution of binary systems
\citep{1964PhRv..136.1224P}, and the detectability horizon of stellar-remnants
as GW sources \citep{1967PhRvL..18.1071F} laid the foundation for the field of
GW astronomy as it resides today. However, a complete description for the rates
and characteristics of stellar populations that produce detectable GW signals
remains elusive. This is because such a description relies on a complicated, and
not yet fully solved, process that requires understanding of the evolution of
stellar systems and the formation environments they reside in. As
\cite{1967PhRvL..18.1071F} discuss in their manuscript: ``Although there are
roughly $10^8$ observable stellar systems within $3000$ light years, of which
approximately $10^5$ are binary systems with periods less than a day, no neutron
star has yet been identified, much less a neutron star binary system; therefore,
it is not possible at the present time to estimate the frequency of occurrence
of such an event, except to say that \emph{it is probably low}\footnote{emphasis
made by author}." Today, population synthesis of GW sources remains a highly
active area of research where new developments range from computational and
statistical techniques, to advances in the understanding and consideration of
environmental effects on the production of GW sources, to the inclusion of new
theoretical or empirically motivated models for the production of all kinds of
GW sources. 

One of the most compelling benefits of detecting close stellar remnant
populations with GWs is the fact that such populations largely remain elusive in
EM surveys. Lacking observations of large stellar remnant populations directly
motivates the need for population synthesis predictions which combine
theoretical predictions for the evolution of individual stellar systems of
interest with the effects of astrophysical environments which shape the age,
composition, spatial distribution, and dynamical influence of each population.
The following sections of this chapter consider first a historical record of the
first predictions of source populations for both ground- and space-based GW
observatories. We then describe the techniques that are employed to carry out
population synthesis studies in a variety of astrophysical environments that can
produce and influence GW source populations. We detail predictions of both the
rates and features of source populations observable with ground and space-based
GW detectors and finish with discussion of complementary source populations from
EM surveys that can aid in the interpretation of current and future GW
population catalogs.

\section{Early, empirically motivated predictions of GW source
populations}\label{chap5:sec2} 

The earliest predictions of GW source populations originated for the so-called W
Ursa Majoris (W UMa) binary population consisting of pairs of low mass
($M\lesssim 1\,\rm{M_{\odot}}$, F/G/K spectral type) stars with close orbits
($0.25\,\rm{day} \lesssim P_{\rm{orb}}\lesssim 1\,\rm{day}$) such that
\emph{both} binary components fill their Roche lobes
\citep{1966SvA.....9..752M}. This `(over)contact' orbital configuration is
stable on Gyr timescales and evolves primarily due to angular momentum loss
driven by magnetic braking effects of the rapidly rotating, magnetized stellar
components rather than GW emission \citep{1981ApJ...245..650M}. Shortly after W
UMa predictions were made, close double white dwarfs (WD-WDs) were also
recognized as potential gravitational wave sources \citep{1967AcA....17..287P}.

The discovery of the Hulse Taylor Pulsar in 1974 \citep{1975ApJ...195L..51H}
ignited community-wide interested in double neutron star (NS-NS) systems. The
first predictions for the birth rate of close/merging double compact object
(DCO) systems containing NS or black hole (BH) components followed shortly
after, building on theoretical work to describe the evolution of close binary
systems in other contexts like the Galactic X-ray binaries
\citep[e.g.][]{1971ARA&A...9..183P, 1975ApJ...198L.109V}. By considering the
volumetric rate of $3$ X-ray binaries within $3\,\rm{kpc}$ and an X-ray lifetime
of $5\times10^4\,\rm{yr}$, \cite{1976ApJ...210..549L} argued for an XRB birth
rate of $2\times10^{-12}\,\rm{pc^{-3}yr^{-1}}$. Assuming that a fraction,
$\beta$, of XRBs survive the explosion of the secondary companion,
\cite{1977ApJ...215..311C} made the first prediction of the rate of close DCO
binaries 
\begin{equation}
  \frac{dN}{dt} \sim 2\times10^{-12}\,\rm{pc^{-3}yr^{-1}} \times \frac{5\times10^4\,\rm{yr}}{1\times10^{10}\,\rm{yr}} \times {\pi\times(100\,\rm{kpc}^2)} \sim 6\times10^{-6}\frac{\beta}{10^{-2}}\rm{yr}^{-1}
\end{equation}

\noindent Based on this rate calculation, under the assumption of
$\beta=10^{-2}$, \cite{1977ApJ...215..311C} concluded that DCO binary rates are
likely to be rare, and less preferred as a GW source population by number when
compared to core-collapse supernovae. 

Between the predictions of \cite{1976ApJ...210..549L} and
\cite{1977ApJ...215..311C} and the advent of computational algorithms to
determine the rates and characteristics of GW sources of close double stellar
remnant binaries, several analytic population synthesis calculations motivated
the development and implementation of plans for ground- and space-based
gravitational-wave detectors. Each of these relied on mostly analytic
calculations \citep[e.g.][]{1973NInfo..27...70T, 1976IAUS...73...35V} which were
later implemented in computational algorithms as computational resources became
more widely available \citep[e.g.][]{1983SvA....27..163K,1987SvA....31..228L}.

The first comprehensive GW population synthesis studies focused on the mHz
frequency regime where close binary stellar remnant systems were expected to
reside despite no known individually resolvable sources at the time
\citep[e.g.][]{1987SvA....31..228L,1990ApJ...360...75H}. These studies relied on
models informed by empirical results for the space density and properties of
observed populations including low-mass binaries still on the main sequence
(often called unevolved binaries), mass transferring systems like W UMa and
cataclysmic variables, in addition to close double stellar remnant pairs
containing WDs, NSs, and BHs. By applying simplified assumptions of constant
star formation over several Gyr, an assumption of GW emission driving the
evolution of double stellar remnants, and a double exponential disk spatial
distribution with $\rho(R,z)\propto \exp{-R/R_0} \exp{-z/Z_0}$, these studies
showed that that the Galactic stellar population likely hosts a rich combination
of GW sources that will produce signals that overlap in frequency space below 1
mHz. 

At the same time that early population synthesis calculations were maturing,
computational algorithms to simulate stellar evolution according to the time
evolving stellar structure equations were also being developed
\citep[e.g.][]{1970AcA....20...47P,1970AcA....20..195P,1971AcA....21....1P,1971AcA....21..271P,1971MNRAS.151..351E,
1972MNRAS.156..361E}. As the stellar evolution codes and theory for close binary
evolution matured
\citep[e.g.][]{1976ApJ...209..829W,1976ApJS...32..583W,1977ApJ...211..486W}, so
did the population synthesis calculations. By the turn of the 21st century, GW
population synthesis studies incorporated sophisticated calculations for effects
of single star evolution implemented with fitting formulae
\citep{1989ApJ...347..998E}, binary star interactions
\citep{1971ARA&A...9..183P} and observational biases which affect assumptions
for the zero age main sequence state of the binaries \citep{Tutukov1980}.

During the late 1990's several themes emerged for predictions of GW source
populations containing NS and BH populations. The ability to detect DCOs with BH
components at much larger distances than those containing only NSs forecasted
that BH-hosting binaries should be detected at larger rates
\citep[e.g.][]{1997MNRAS.288..245L,1998ApJ...506..780B,2003MNRAS.342.1169V}. It
was also formalized that both Roche-lobe-overflow interactions and supernova
explosions leading to CO natal kicks play a critical role in the formation and
merger of DCO populations
\citep[e.g.][]{1996A&A...309..179P,1998A&A...332..173P,1999A&A...346...91B}.
Selection effects, and the difficulty of calculating them for Galactic
populations, were incorporated into rate predictions
\citep{1991ApJ...379L..17N,2000ApJ...530..890K,2003ApJ...584..985K} and jointly
considered with the effects of binary-star interactions \citep[e.g.][and
references therein]{2001ApJ...556..340K}. It was also recognized that the
inclusion of the effects of dynamically active environments were critical to
provide the most complete set of predictions for GW source populations. As a
result, binary population synthesis tools were developed with the direct aim of
including them in direct Nbody calculations \citep[e.g.][]{1996A&A...312..670P,
2001MNRAS.323..630H}. These early population synthesis predictions paved the way
for the construction and support of ground-based GW detectors like the Laser
Interferometer Gravitational-Wave Observatory (LIGO), Virgo, and KAGRA.

Similar themes emerged for predictions of GW source populations in the mHz
frequency band. It was shown that the Galactic population of WD-WD binaries will
be the predominant source of GWs in the mHz regime and will vastly outnumber the
BH and NS hosting binary systems
\citep{1990ApJ...360...75H,1995A&A...298..677L}. These predictions led to
estimates for the so-called confusion foreground of GWs caused by overlapping
signals from WD-WD binaries across the Galaxy, with refinements being
incorporated from observational limits placed by Type Ia supernova rates
\citep{1998ApJ...494..674P}, updates in population synthesis calculations for
Galactic populations \citep{Nelemans2001,Nelemans2001b}, and considerations of
GW backgrounds from cosmological populations of stellar-origin sources
\citep{2001MNRAS.324..797S,2003MNRAS.346.1197F}. Individual sources that could
be potentially resolved by LISA were also considered in the context of
multimessenger observation opportunities for both detached and accreting WD-WD
binaries \citep{2004MNRAS.349..181N, 2005ApJ...633L..33S} as well as probes of
the dynamical stellar populations of the Milky Way globular cluster population
\citep{2003MNRAS.346.1197F}. Similar to the development of ground-based
detectors, these early studies played a central role in the proposal and
eventual the adoption of the Laser Interferometer Space Antenna
\citep{2017arXiv170200786A} scheduled to launch in the mid 2030s. 

\section{Population synthesis tools and techniques}\label{chap5:sec3} 

The importance of the effect of astrophysical environments on the production of
the GW sources they host was recognized early on in the development of
modern-day population synthesis tools. Since the vast majority of proposed GW
formation scenarios involve a stellar origin, and since stars are observed to
form in a variety of environments, it is natural that the tools needed to
simulate stellar populations and the GW sources they produce followed suit.

\begin{figure}
  \centering
  \includegraphics[width=0.75\textwidth]{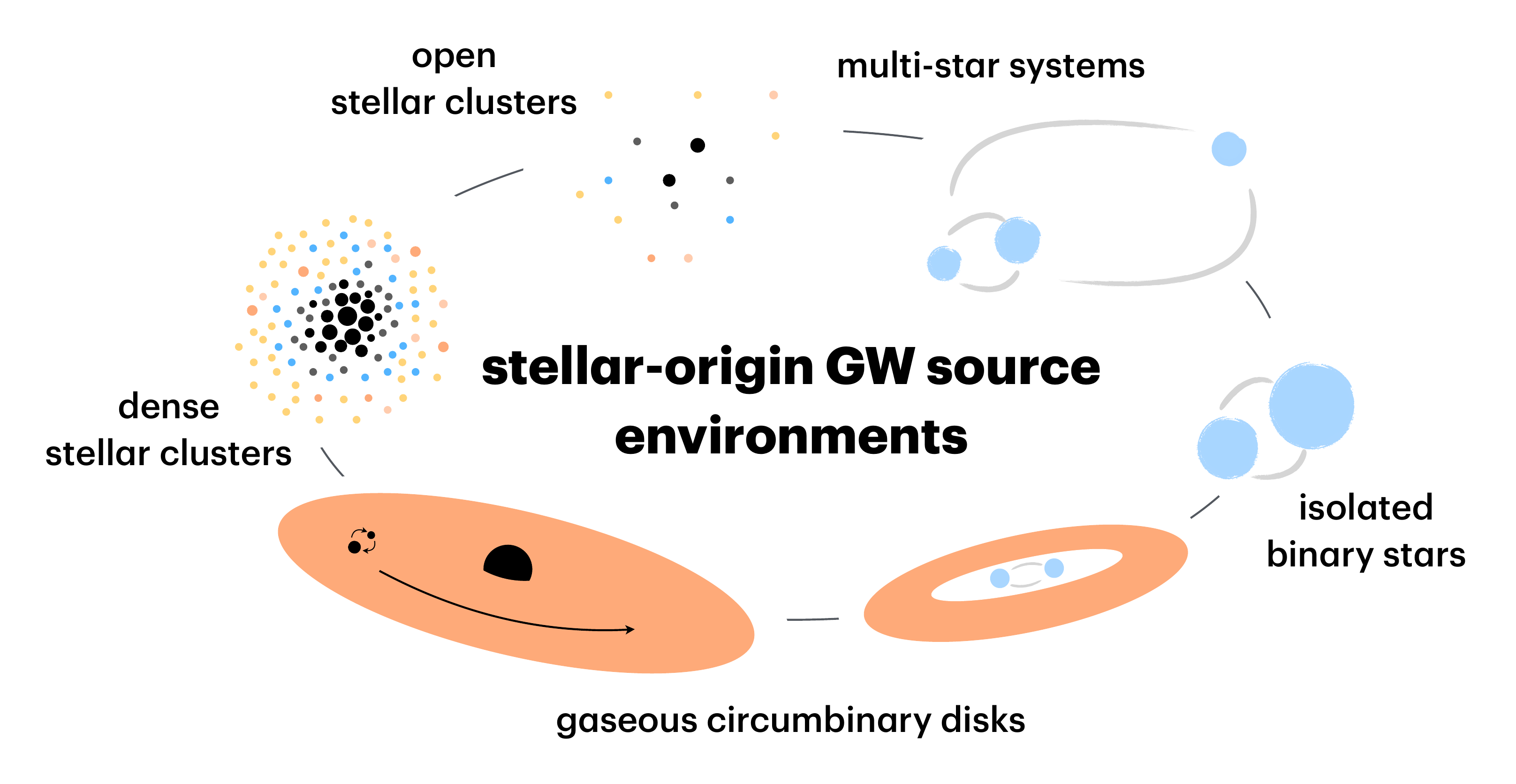}
  
  \caption{A cartoon of the rich variety of the astrophysical environments which
  could host stellar-origin GW sources. This does not include the potential
  formation of binaries consisting of primordial BHs formed from early Universe
  fluctuations.}\label{fig:channels}
\end{figure}

Following the momentous discovery of the first GW source, GW150914
\citep{150914}, work quickly emerged showing that GW source populations could
originate in an even wider variety of astrophysical environments than previously
considered. In addition to rate estimates for merging pairs of DCOs originating
in isolated binaries \citep[][]{Belczynski2016} or globular clusters
\citep[][]{Rodriguez2015}, rate estimates emerged for triple systems residing in
the globular clusters \citep{Antonini2016} and the Galaxy \citep{Antonini2017},
nuclear star clusters \citep{Antonini2016b}, the disks surrounding active
galactic nuclei (AGN) \citep{Bartos2017}, and even primordial BHs originating
from early Universe fluctuations \citep{Bird2016}. The subsequent releases of
the first, second and third GW transient catalogs \citep[GWTC-1, GWTC-2, and
GWTC-3][]{GWTC-1, GWTC-2, GWTC3} have fueled the refinement and further
development in each channel in addition to the recent consideration of wide
binaries driven to merger by fly-by interactions \citep{Raveh2022, Michaely2022}
and circumbinary disks which could be formed through Roche-lobe-overflow binary
interactions that lead to hydrogen-rich material leaving the binary system, but
still driving orbital evolution \citep[e.g.][]{Siwek2023a}.
Figure~\ref{fig:channels} shows a cartoon version of the rich ecosystem of
astrophysical environments that can produce GW sources. 

To cover the wide array of potential GW source hosts, there are several
population synthesis tools presently in use by the astronomical community which
span several different astrophysical environments. Many of these tools are
dependencies of other population synthesis software, thus forming a rich
development ecosystem bolstered by the open collaboration of several groups
across the globe. In the following sections, we discuss the landscape of
population synthesis software used to simulate GW source populations originating
in each environment. We briefly summarize and discuss the costs and benefits of
each below.

\subsection{Isolated binary evolution}\label{chap5:sec3subsec1} 

The simplest population synthesis case is that of single star populations.
However, a growing body of observations has shown that \emph{most} stars form
with a companion \citep[][]{Raghavan2010,Sana2012,Moe2017,Offner2023}. For
lower-mass stars ($M\lesssim5\,\rm{M_{\odot}}$) the fraction of close binaries,
with orbital periods below $10,000$ days, decreases with increasing metallicity
\citep{Moe2019}. While for massive stars, the close binary fraction remains
large and the \emph{companion} fraction increases beyond unity, indicating that
many massive stars reside in higher-multiple systems \citep{Moe2017}. Thus, for
any stellar populations that reside in galactic fields and therefore do not
undergo significant dynamical interactions, it is critical to consider the
impact of binary-star interactions. In doing so, most binary population
synthesis techniques can incorporate single-star populations with little to no
effort. In this subsection we consider three types of binary population
synthesis codes which vary primarily in level of detail that stellar evolution
is treated in the presence of binary-star interactions. 

\subsubsection{Rapid codes}\label{chap5:sec3subsubsec1} The largest class of
population synthesis tools used to simulate populations of isolated binaries are
the so-called `rapid' codes which apply fitting formulae to a grid of single
stars run with a single stellar evolution model. The vast majority of rapid code
bases are built using a series of algebraic fitting formulae, often referred to
as the `Hurley' fits \citep{2000MNRAS.315..543H} which were applied to the OVS
grid of stellar evolution models published in \cite{1998MNRAS.298..525P}. This
grid was produced using the stellar evolution code initially developed by Peter
Eggleton \citep{1971MNRAS.151..351E, 1972MNRAS.156..361E,1973MNRAS.163..279E}
and subsequently undergoing several updates
\citep{1994MNRAS.270..121H,1995MNRAS.274..964P}. The grid was evolved with
masses ranging from $0.1-50\,\rm{M_{\odot}}$ and with metallicities $Z= 0.0001,
0.0003, 0.001, 0.004, 0.01, 0.02,$ and $0.030$ where the hydrogen and helium
abundances are defined as functions of the metallicity $X=0.76-3.0Z$ and $Y =
0.24+2.0Z$. These fits were first implemented in the Single Star Evolution (SSE)
algorithm and shortly after, a Binary Star Evolution (BSE) algorithm built on
the fits that incorporated several effects of binary-star interactions was
released \citep{2002MNRAS.329..897H}. 

Since the release of BSE, several rapid codes have been developed based on the
Hurley tracks, and in some cases, the BSE algorithm. Each code is developed with
different use cases in mind are thus implemented across several languages and
employ various updates to binary evolution prescriptions. Codes in use today
that are based on the Hurley tracks include SeBa
\citep{1996A&A...309..179P,2012A&A...546A..70T}, StarTrack
\citep{2008ApJS..174..223B}, binary\_c \citep{Izzard2004, Izzard2023} and
binary\_c-python \citep{Hendriks2023}, COSMIC \citep{Breivik2020a}, COMPAS
\citep{2022ApJS..258...34R}, and MOBSE \citep{2018MNRAS.480.2011G}. Scenario
Machine is another rapid code that is not based on the Hurley tracks but
incorporates the effects of single and binary star interactions in a consistent
way to other rapid codes listed above. 

\begin{figure}
  \centering
  \includegraphics[width=0.65\textwidth]{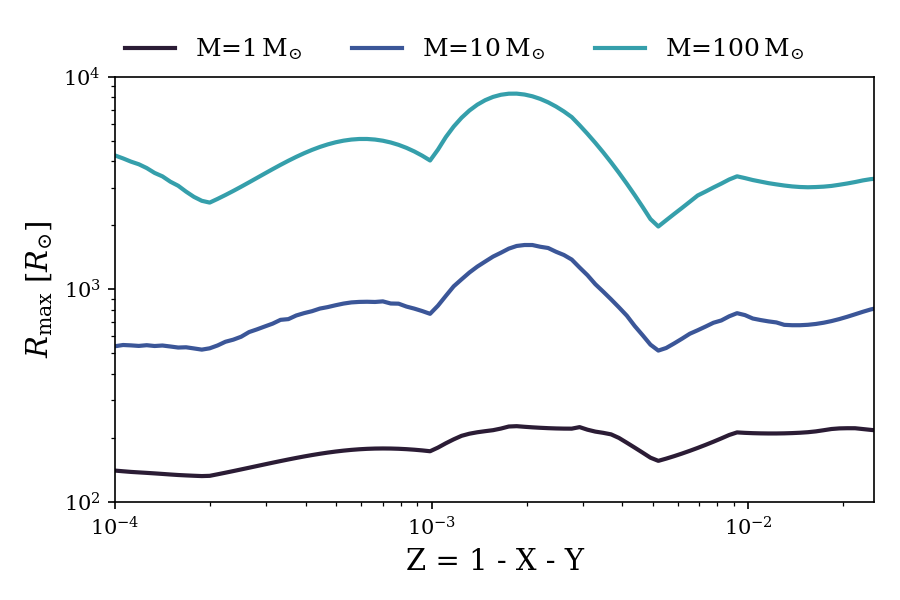}
  
  \caption{The maximum radius of a single star plotted as a function of the
  metallicity of the star where each color shows a different mass. The radius
  behavior is an unphysical artifact induced from the fitting formulae of
  \cite{2000MNRAS.315..543H} that are implemented in most rapid binary
  population synthesis codes.}\label{fig:rapid-radius}

\end{figure}

Rapid codes are widely used to make predictions for GW sources of all stellar
remnant types. A major advantage in their use is the ability to simulate large
binary populations that span orders of magnitude in metallicity and cover a
large swath of uncertain models for binary interactions. Since they rely on fits
to single star evolution, they are unable to capture any effects of uncertain
stellar physics like the impacts of changing convection assumptions, nuclear
reaction rates, or including rotation. They are also often used to extrapolate
far beyond the mass ranges of the original OVS tracks. This is the case for
nearly all BH-BH binary mergers which originate from stars often significantly
more massive than $50\,\rm{M_{\odot}}$. 

Finally, due to the enormous range in mass and metallicity space that the
fitting formulae must cover, there are regions of the predictive parameter space
where the fits are not always adequate. One such example is the maximum radius
of a star with fixed mass but changing metallicities. In this case, the maximum
radius would be expected to be a continuous function that responds to the
effects of metallicity-specific wind mass loss. Figure~\ref{fig:rapid-radius}
shows the actual behavior of the maximum stellar radius as a function of
metallicity for a variety of masses as implemented in COSMIC. Since the majority
of GW progenitors will interact well before the maximum stellar radius is
achieved, this effect is unlikely to impart uncertainties at the level of
previously discussed uncertainties. Nevertheless, it is a virtuous scientific
endeavour to reduce computational modeling artifacts as much as possible.

\subsubsection{Hybrid codes}
Hybrid codes offer an additional level of complexity beyond the rapid codes
because they include interpolated tracks from single star evolution simulations
that can account for how rotation and convection affect populations in addition
to providing information on the structure of each star. There are three hybrid
codes in use at present: SEVN \citep{2017MNRAS.470.4739S,2023MNRAS.524..426I},
ComBinE \citep{2018MNRAS.481.1908K}, and METISSE \citep{2020MNRAS.497.4549A}.
ComBinE uses a densely sampled grid of stellar models evolved by the Bonn
Evolutionary code (BEC) and builds on the work of \cite{2010ApJ...725..940Y}.
METISSE and SEVN are engineered to be used agnostically with single star grids
as long as they follow the equivalent evolutionary phase (EEP) data format
described by \cite{2016ApJS..222....8D}. EEPs were first implemented for MESA
simulations to create the MESA Isochrones and Stellar Tracks (MIST) grid of
single stars and their isochrones, though the concept of EEPs dates back to much
earlier work \citep[e.g.][]{1970ApJ...159..895S}. So far, SEVN has been
developed to work with with both MIST and PARSEC single star grids while METISSE
is compatible with grids from MIST, BEC, and the Cambridge ev code. 

All three hybrid codes allow for the incorporation of information produced by
single stellar evolution codes that are not traditionally included in the
fitting formulae used in the rapid codes. These include chemical yields (note
that binary\_c incorporates these as well), radial density profiles, and
late-stage nuclear burning phases. The treatment of wind mass loss is usually
handled at the grid simulation stage such that winds are self-consistently
applied with the interior evolution of both stars in the binary. However, the
incorporation of binary star interactions is applied in the same way as the
rapid codes. This allows for direct comparisons to current and previous results
from rapid population synthesis studies but also means that hybrid codes are
unable to resolve the effects of Roche-lobe overflow mass transfer on neither
the donor nor the accretor's interior evolution. 

\subsubsection{Detailed codes}
The use of detailed stellar evolution codes, where both stars in a binary are
evolved simultaneously, is the most computationally taxing technique for
population synthesis of isolated binary star GW source progenitors. However, it
is also the most accurate application of the stellar structure equations.
Several studies have been used to predict GW populations with detailed stellar
evolution modeling
\citep[e.g.][]{2017A&A...604A..55M,2018A&A...616A..28Q,2020A&A...637A...6L},
however each of these studies were done with specific aims that often cover a
restricted binary-star parameter space. Two population synthesis codes based on
detailed binary evolution modeling are currently used for general purpose
predictions of GW source populations: BPASS \citep{2017PASA...34...58E,
2018MNRAS.479...75S} and POSYDON \citep{2023ApJS..264...45F}.

The present version of BPASS (detailed in \cite{2022MNRAS.512.5329B}) is based
on a modified version of the Cambridge STARS code and does not compute the
evolution of each star in a binary simultaneously. Instead, the Hurley tracks
are used for the evolution of the less massive star (or secondary) until the
more massive star (the primary) evolves to become a stellar remnant. The
secondary's evolution is then computed with accretion and rejuvenation
calculated according to the mass transfer that is expected to occur during the
primary star's evolution. POSYDON, on the other hand, computes the simultaneous
evolution of each star using MESA through a combination of single star evolution
tracks stored as EEPs and interacting binary-star grids that can be interpolated
between for use in population synthesis studies. 

While both codes are able to incorporate the effects of Roche-lobe overflow
accretion in the interior evolution of stars, assumptions must be made for how
the mass transfer proceeds including the rate at which mass leaves the donor
star, the fraction of mass that is accreted by the companion, and how mass that
is not accreted leaves the system. Given the already immense computation costs
of these detailed calculations, both BPASS and POSYDON only allow for the
self-consistent application of a single set of Roche-lobe overflow assumptions.
Applications of CO formation models can, however, be widely applied in a
post-processing fashion such that different assumptions for natal kick and
supernova physics can be applied within the already computed grids for both
BPASS and POSYDON.

\subsection{Triples and higher multiples}\label{chap5:sec3subsec2} 

In this subsection, we consider the addition of an outer, tertiary companion to
an inner binary system. While the addition of a single body may sound
inconsequential at first, the evolution of a triple system represents an
enormous amount of complexity to consider both on the orbital dynamics of the
system as well as the potential influence a triple companion may have on how
stellar evolution proceeds for each component of an inner binary. Nevertheless,
when considering the formation of GW sources, we cannot ignore the possibility
of a non-negligible portion of the observed populations originating in triple
systems. Indeed, we remind the reader that a significant fraction of massive
stars are observed to have \emph{at least} one companion \citep{Moe2017,
Offner2023}. 

There are four codes recently or currently in use. These all employ rapid binary
population synthesis codes to evolve stars in the context of triple star
systems: TRIPLE\_C (which builds on binary\_c, \cite{2013MNRAS.430.2262H}), TRES
(which builds on SeBa \cite{2016ComAC...3....6T}), MSE (which builds on
BSE\cite{2021MNRAS.502.4479H}), and TSE (which builds on MOBSE
\citep{2022MNRAS.516.1406S}). Other codes have been developed to simulate the
secular effects of triple dynamics but do not include the impact of binary-star
interactions: KOZAI \citep{Antognini2015}, KOZAI (identically named)
\citep{Antonini2018}, and SECULARMULTIPLE \citep[which is not restricted to only
three bodies,][]{Hamers2016, Hamers2018, Hamers2020}.

\subsection{Stellar clusters}\label{chap5:sec3subsec3} 

Recognition of stellar clusters, and more specifically globular clusters, as
potential hosts for GW sources was attained nearly as early as recognition for
isolated binaries as progenitors of isolated binaries
\citep[e.g.][]{PortegiesZwart2000}. Indeed, the development of rapid algorithms
that allow for stellar evolution to be considered in stellar cluster
environments played a central role in the development of the rapid codes used
for isolated binary evolution in use today (c.f.
Section~\ref{chap5:sec3subsubsec1}). The simulation of stellar clusters is
divided into multiple classes: those carrying out the direct summation of all
interactions within a system of N bodies (the so-called N-body approach), those
employing orbit-averaged Monte Carlo methods and those which use further
assumptions which simplify the calculation process beyond the Monte Carlo
method. 

Several codes have been developed to carry out direct N-body calculations, many
of which employ 4th order Hermite integrators, including PhiGRAPE
\citep{Harfst2008}, ph4 \citep{McMillan2012}, HiGPUs
\citep{CapuzzoDolcetta2013}, frost \citep{Rantala2021}, PeTar \citep{PETAR}, and
the well-known NBODY series \citep{Aarseth2003, Aarseth2012} and its GPU
implementation \citep{Wang2015}. These codes calculate the gravitational force
between every pair of particles and sum up the effects of each pair to compute
the instantaneous acceleration of each body in the system. They typically
include stellar and binary evolution physics based on SSE and BSE (these tools
were developed with the express intent for including them in N-body integrators;
\cite{Hurley2001}). Despite their accuracy, direct summation methods are
exceeding computationally expensive, thus limiting their application to star
cluster systems with low binary fractions and large initial radii, both of which
experience fewer dynamical interactions \citep{Heggie2014, Wang2016,
Rantala2021}. 

The orbit-averaged Monte Carlo, approach pioneered by Hénon
\citep{1971Ap&SS..13..284H, 1971Ap&SS..14..151H}, offers a less computationally
expensive alternative by statistically modeling stellar encounters such that the
cumulative effect of many distant two-body encounters is modeled as a single
effective scattering between neighboring particles underlying the assumption of
a spherically symmetric cluster. The simplified modeling in Hénon's Monte-Carlo
method naturally resolves the two-body relaxation that drives the secular
evolution of collisional star systems and has been shown to reproduce the
evolution of dynamically active star clusters in several studies spanning over
five decades \citep[e.g.][]{1974A&A....37..183A,Joshi2000,Kremer2020}. This
method has been implemented in two widely used codes: CMC \citep[which employs
COSMIC for the treatement binary evolution; ][]{Pattabiraman2013,Rodriguez2022}
and MOCCA \citep[which employs BSE; ][]{Giersz2008,Giersz2013}. Comparisons
between direct summation and Monte Carlo approaches show that simulated star
clusters with order $10^6$ systems agree well \citep{Rodriguez2016a}, though as
the number of stars in the cluster decreases below $\sim10^4$ the Monte Carlo
approach begins to break down due to a departure from the assumed spherical
symmetry of the method. In this case, direct summation methods are required
\citep[e.g][]{Banerjee2017, DiCarlo2019}. 

For larger clusters with population numbers exceeding $10^6$ stars, common for
nuclear star clusters found at the centers of galaxies, Monte Carlo approaches
begin to become computationally infeasible. In these cases, further simplifying
assumptions can be applied to reduce computational costs. For example, building
on the well-understood impact of mass segregation in large clusters,
\cite{Antonini2016b} use the BH mass functions found from simulations with CMC
\citep{Rodriguez2015} to then perform semianalytic calculations for how a
nuclear star cluster would produce BH-BH mergers. More recently, codes like
FASTCLUSTER \citep{Mapelli2021,Mapelli2022} and Rapster \citep{Kritos2024} have
been used to apply semianalytic functions for the evolution of star clusters
under the assumption of spherical symmetry. 

\subsection{Gaseous environments: Active galactic nuclei and circumbinary
disks}\label{chap5:sec3subsec4} 

The incorporation of the effects of gaseous environments on the production of GW
sources is relatively recent in comparison with the consideration of other
astrophysical environments. As such, the implementation of software that
generates population predictions for stellar populations that consider the
presence of a circumbinary disk around individual binaries or stellar
populations residing in gaseous disks associated with active galactic nuclei
(AGN) remains less broad than other environments. This is compounded by the
complicated interactions with the disk that must be accounted for that are
driven by the wide variety of potential binary mass ratios, orbital separations,
orbital eccentricities (both within the star and potentially around an AGN disk)
and inclinations with respect to either a circubminary or AGN disk. Yet another
level of compounded complication is the changes to stellar evolution and binary
interactions that could be caused through interactions with a disk. 

The incorporation of the effects of circumbinary stellar disks on the production
of GW sources has generally been considered for specific test cases due to
either theoretical models tailored to specific observed phenomena
\citep[e.g.][]{Tuna2023} or simulation-based studies which rely on the necessary
inclusion of hydrodynamical simulations like Arepo
\citep{Springel2010,Pakmor2016} or DISCO \citep{Duffell2016} to account for the
evolution of the disk \citep[e.g.][]{Siwek2023a, Siwek2023}. On larger scales,
the effects of an AGN disk on the production of GW sources has been studied with
varied methods including semianalytic models
\citep[e.g.][]{Stone2017,McKernan2020,Tagawa2021}, hydrodynamical modeling
\citep[e.g.][]{Li2021,Dittmann2024} with codes like Athena++ \citep{Stone2020}
or LA-COMPASS \citep{Li2005}, stellar evolutionary modeling
\citep{Cantiello2021}, and direct summation dynamical modeling which includes
the effects of a central supermassive black hole \citep[e.g.][]{Secunda2019}. At
present there are two published population synthesis tool to simulate GW
populations across a variety of assumptions for AGN disks: starsam, a
semianalytic model that accounts for stellar population evolution in an AGN disk
through hydrogen burning \citep{Dittmann2024b} and McFACTS which incorporates
the effects of both dynamical encounters and migration within a disk for compact
object populations \citep{McKernan2024McFACTS, 2024arXiv241110590C,
Delfavero2024McFACTS}.

\section{Predicted rates of GW sources across astrophysical environments}
\begin{figure}
  \centering
  \includegraphics[width=1\textwidth]{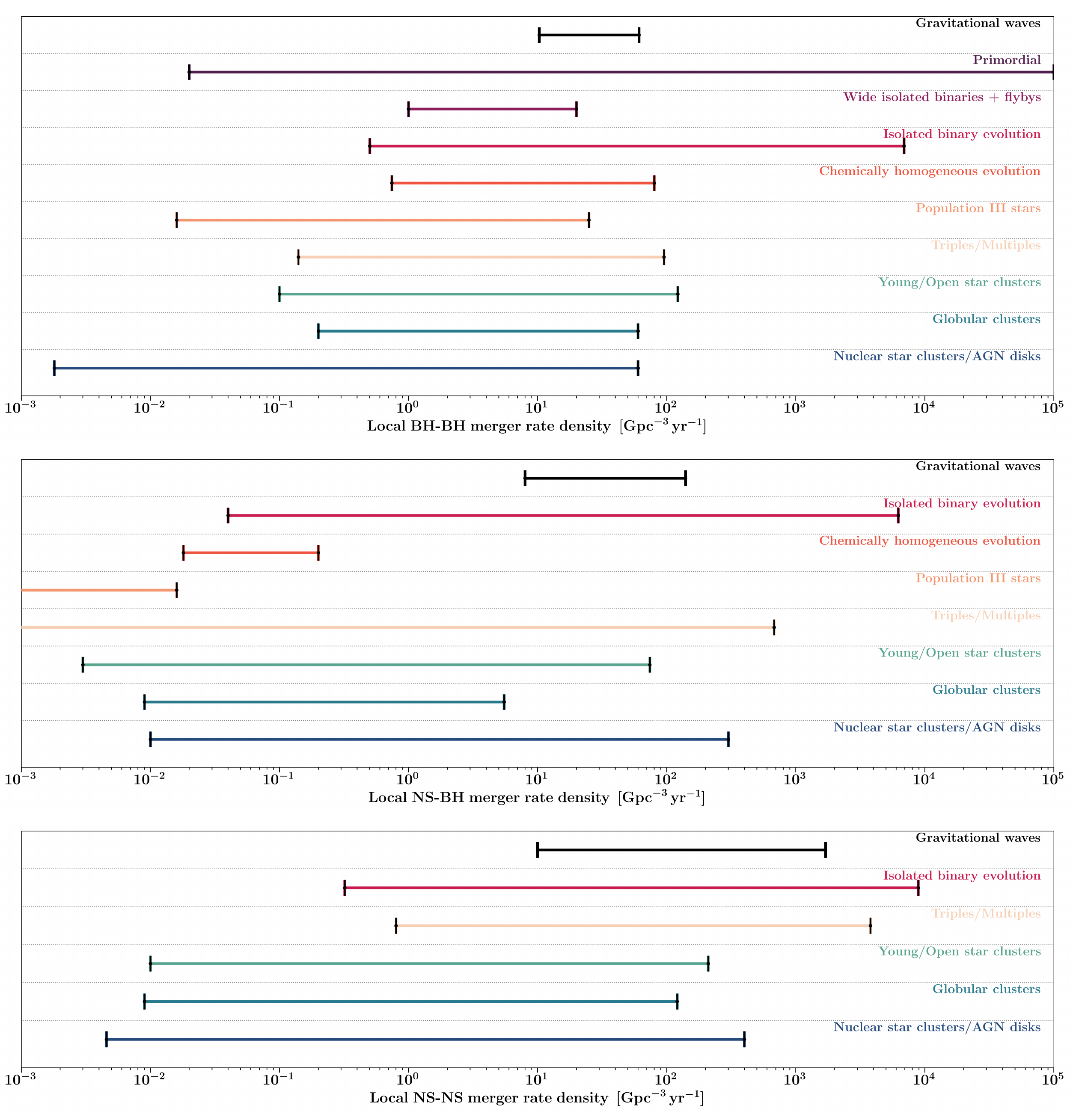}
  
  \caption{Predicted merger rates for different combinations of NS and BH
  hosting GW sources originating in different astrophysical environments.
  Adapted from \cite{Mandel2022} and \cite{ZenodoReview:2021}; we refer the
  reader to the in depth discussion of DCO merger rates discussed in
  \cite{Mandel2022}.}\label{fig:dco-rates}
\end{figure}

The primary use of population synthesis software in the context of GW sources
has been to predict the occurrence rates of sources for both ground and
space-based GW detectors. As such, the vast majority of results published in
papers using population synthesis tools are aimed at calculating either the
local merger rates of DCO binaries or the local population of binaries hosting
WDs or COs radiating GWs with mHz frequencies. Figure~\ref{fig:dco-rates},
adapted from \cite{ZenodoReview:2021}, shows the limits of rates that have been
predicted for DCO binaries originating in a wide variety of astrophysical
environments. We briefly discuss the process for calculating these rates and the
quoted ranges for each environment but encourage the reader to refer to
\cite{Mandel2022} which discusses the studies used to make the figure in detail.

\subsection{Local merger rates for double compact object binaries}
The calculation of local merger rates of DCO binaries is usually done in two
parts. First, the merging DCO population is simulated using a population
synthesis tool designed for the environment of interest. It is often the case
that several models are considered for uncertain physics that govern the
production of GW sources (e.g. the interactions between stars in a binary;
\cite{Broekgaarden2022} or the initial conditions for globular cluster
formation; \cite{Kremer2020}). These simulations usually result in a synthetic
catalog of DCO mergers with a range of metallicity (either sampled continuously
or on a grid) and merger properties, including the time from the formation of
the CO progenitors in each environment. Note that in the case of primordial BHs,
this formation time occurs in the early Universe \citep{Bird2016, Ali-Haimoud},
while for stellar-origin COs, the formation time is usually taken to be the time
of the stellar zero-age main sequence. 

After the simulation of the synthetic merger catalog, a cosmic star formation
history that accounts for the formation of all DCO populations within a Hubble
time is assumed and applied through a rate equation, often of the form

\begin{equation}
  R_{\text{DCO}}(z) = \int \mathrm{d}Z \int_0^{t_m(z)} \mathrm{d}t'_{\text{delay}} 
    \times \frac{\mathrm{d}^2 N_{\text{form}}}{\mathrm{d}M \mathrm{d}t_{\text{delay}}}(Z, t'_{\text{delay}})
    \times S(Z, z(t_{\text{form}})).
\end{equation}

\noindent Here, the rate, $R_{\text{DCO}}(z)$, is the merger rate of a given DCO
in the source frame at redshift $z$. The integrals over the metallicity, $Z$,
and delay times, $t'_{\text{delay}}$, are designed to account for all DCO
mergers that could occur up to merger time, $t_m(z)$, given their formation
time, $t_{\text{form}}$, as well as their metallicity. The first term in the
integral represents the number of mergers formed per unit mass per unit delay
time, and depending on the environment may be cast in terms of the number of
mergers per cluster or number of mergers per AGN disk instead. The second term
in the integral represents the amount of stars formed per unit metallicity at
formation redshift, $z(t_{\text{form}})$, in the case of isolated binaries. For
other environments, this term may represent the formation rate of globular
clusters or AGN at a given metallicity and formation redshift.

The most widely applied star formation history models are empirical ones based
on observations from spectroscopic and photometric \citep[e.g.][]{Madau2014,
Harris2013} or X-ray surveys \citep[e.g.][]{Madau2017}. Another option is to use
cosmological simulations that account for star formation
\citep[e.g.][]{Mapelli2019}, globular clusters \citep[e.g.][]{Rodriguez2023}, or
nuclear star clusters \citep[e.g.][]{Barausse2012} across cosmic scales. The
incorporation of stellar population formation adds an extra layer of uncertainty
to each rate calculation beyond the uncertainties for the formation of DCO
mergers within each environment \citep[e.g.][]{Chruslinska2019, vanSon2023,
Chruslinska2024}. Indeed, for the case of BH-BH mergers formed through isolated
binary evolution, the uncertainties for the metallicity-specific cosmic star
formation rate are larger than the uncertainties within the binary evolution
modeling \citep{Broekgaarden2022}. This is largely due to the strong metallicity
dependence of wind mass loss for massive stars in rapid population synthesis
modeling following the empirical studies from \cite{Vink2001} and
\cite{Vink2005}.

\subsubsection{Primordial black holes}
The widest uncertainties reported for local DCO merger rates belong to
primordial BH binaries, with \cite{Ali-Haimoud} reporting the widest merger rate
range. They consider the effects of tidal torquing due to all primordial BH
binaries, which depends heavily on the fraction of non-relativistic dark matter
and baryons that primordial BHs account for. As the fraction increases from one
part in $10^4$ up to $1$, the local merger rate increases from
$0.2\,\rm{Gpc}^{-3}\rm{yr}^{-1}$ to $10^5\,\rm{Gpc}^{-3}\rm{yr}^{-1}$. The lower
limit from \cite{Ali-Haimoud} is increased from the lower limit of
$0.02\,\rm{Gpc}^{-3}\rm{yr}^{-1}$ reported in \cite{Bird2016} due to the
inclusion of tidal torques which shortens merger times for primordial BH-BH
binaries. For an in-depth discussion of primordial black holes, see their
dedicated chapter as part of this encyclopedia. 

\subsubsection{Isolated binaries}\label{subsubsec:isolated-mergers} 

Several sub-channels exist for DCO mergers that originate in binary systems in
isolation, typically those residing in galactic environments outside of stellar
clusters. At the widest scales, binary systems which are not expected to undergo
any mass transfer may still produce DCO mergers through flyby interactions from
stars \citep{Raveh2022, Michaely2022} that are potentially influenced by the
galactic tides \citep{Stegmann2024}. The dominating uncertainty in the rate of
flyby-induced DCO mergers is the strength of natal kicks at the formation of the
CO, with larger kicks unbinding the majority of wide binaries. For the "isolated
binary evolution" channel, which contains standard massive stars in typical
orbits \citep[e.g.][]{Sana2012}, the dominating binary evolution uncertainties
are Roche-lobe-overflow mass transfer and its stability, with natal kick
strengths also playing an important role. In the case of mass transfer
stability, an increased fraction of Roche overflow events leading to stable mass
transfer can significantly decrease the BH-BH merger rate
\citep[][]{Gallegos-Garcia2021}, while an increased common envelope ejection
efficiency can significantly increase the NS-NS population
\citep[][]{Mapelli2019}. 

\subsubsection{Chemically homegenous isolated binaries}
The chemically homogeneous channel only applies to massive stars ($M \gtrsim
30\,\rm{M_{\odot}}$) in very short period binaries ($P \lesssim 2.5\,\rm{hr}$)
with near equal mass companions \citep{Marchant2016, Mandel2016} and low
metallicities ($Z\lesssim 0.006$; note that this is a loose upper limit). For
binaries with these properties, internal mixing driven by rotation at an
appreciable fraction of the break up rotation causes the stars to forego phases
of evolution where their radii expand and fill their Roche lobes. Given the
restricted parameter region where chemically homogeneous evolution can occur,
the largest uncertainty in the rates of the channel arise from the exact
conditions at which it does occur in nature. Less restrictive ranges on the
allowed metallicities can significantly enhance the merger rate for DCOs
containing BHs that originate through the chemically homogeneous channel. 

\subsubsection{Population III isolated binaries}
The lowest metallicity population of binary stars that could produce DCO mergers
containing BHs are those containing population III stars which are theorized to
be the first stars formed and comprised of entirely hydrogen. Typically, the
evolution of population III stars is implemented by applying the radius
evolution from one-dimensional stellar evolution grids
\citep[e.g.][]{Marigo2001} to rapid population synthesis fitting formulae. The
largest rate for DCO mergers originating from population III binaries are
reported by \cite{Kinugawa2014}, while the lowest rates are reported by
\cite{Belczynski2017}. The major differences in these two studies are the
application of initial conditions and the allowed DOC merger channels including
whether donors crossing the Hertzsprung Gap can successfully eject a common
envelope. \cite{Kinugawa2014} applies standard initial conditions used for
population I/II binaries while \cite{Belczynski2017} discusses separate initial
conditions assumptions that are tailored to extremely low metallicity
conditions. In addition to the initial conditions and evolutionary assumptions,
the amount of star formation originated in population III stars can
significantly impact the DCO merger rates \citep[e.g.][]{Hartwig2016}.  

\subsubsection{Triple systems}
The addition of a third body, almost always in a hierarchical configuration
where the inner binary dynamics are driven by a much wider outer body,
significantly increases the complexity of the production of DCO mergers. Because
of their ubiquity in stellar populations \citep[e.g.][]{Offner2023}, DCO mergers
formed in triple systems are expected to occur with stellar components in both
galactic fields \citep[][]{Silsbee2017, Antonini2017}, as well as dynamically
active environments like globular clusters \citep{Martinez2020} or in nuclear
star clusters where the third body is a galaxy's central supermassive black hole
\citep[e.g.][]{Hoang2018, Wang2021}. 

In nearly all cases, the Lidov-Kozai \citep{Lidov1962, Kozai1962} or eccentric
Kozai-Lidov \citep{Naoz2013} mechanisms cause the inner orbit's argument of
pericenter to oscillate which leads to cycles of exchanges in the eccentricity
and inclination of the inner orbit. These exchanges can produce eccentricities
large enough that the GW merger timescale is significantly shortened. In
addition to the binary interaction uncertainties that isolated binaries are
subject to, the initial conditions and rates of triple systems that remain
dynamically unstable are a key source of uncertainties in the rate of DCO
mergers produced by triples. 

\subsubsection{Young and/or open stellar clusters}
Young stellar clusters with densities significantly lower than the stellar
densities of globular or nuclear star clusters combine several of the
uncertainties discussed above. Simulations of young stellar clusters must
additionally consider the impact of dynamical encounters that cannot be treated
with secular approximations. This makes young stellar clusters some of the most
difficult environments to simulate from a computational standpoint; thus large
parameter sweeps which capture the uncertainties associated with binary
interactions are limited \citep[e.g.][]{DiCarlo2019}. Further uncertainties
arise from the occurrence rates and initial properties of open stellar clusters,
with the initial mass of the cluster playing a critical role in the DCO merger
rates \citep{Torniamenti2022}. 

\subsubsection{Globular clusters}
The DCO merger rate for systems dynamically formed in globular clusters is
dominated by BH-BHs due to the synergy between the dynamical evolution of the
cluster itself and the BH population it harbors \citep[e.g.][]{Morscher2015}.
Broadly, this leads to clusters that have either undergone core collapse and
thus have few BHs in their centers due to dynamical ejections, or
non-core-collapsed clusters which likely contain dozens to hundreds of BHs in
the core depending on the initial mass and compactness of the cluster
\citep[e.g.][]{Kremer2019}. Thus similar to isolated binaries, which are subject
to wide uncertainties in the metallicity-specific cosmic star formation history,
the BH-BH merger rate uncertainty for globular cluster populations is dominated
by the formation history and properties of the clusters themselves
\citep[e.g.][]{Antonini2020}. 

The merger rate for NS hosting binaries originating in globular clusters is
expected to be significantly lower. The dynamical mass segregation that produces
large numbers of dynamically formed BH-BHs necessarily pushes the NS population
to larger radii where the densities (and thus dynamical encounter rates) are
lower. This is compounded by the possibly large natal kicks that can exceed the
low ($\sim30\,\rm{km/s}$) cluster escape speeds and thus eject NSs from the
cluster. It is thus expected that the NS-BH and NS-NS merger populations
originating in globular clusters cannot exclusively make up the observed
populations from ground-based detectors \citep[][]{Ye2020}. 

\subsubsection{Nuclear star clusters}
Nuclear star clusters, which occupy the very inner regions of galactic centers,
are the most massive and densest stellar clusters; they thus have much larger
escape speeds than globular clusters. Their density is further increased in the
absence of a central supermassive BH \citep[e.g.][]{Antonini2017}. The increase
in density can lead to higher rates of dynamical interactions which unbind
dynamically formed binaries before they can merge from GW emission. This effect
is especially true for NS-hosting binaries.  However, the increased density also
leads to more hierarchical mergers between BHs due to fewer dynamical ejections
of merged GW sources which can experience a merger-induced kick for non-aligned
BH spins. This is finally convolved with the occurrence rate of nuclear star
clusters which is much lower than the rate of globular clusters which number in
the several hundreds per galaxy \citep{Harris2013}. As such, nuclear star
clusters are predicted to produce DCO merger populations at lower rates than
their globular cluster counterparts. 

\subsubsection{Gaseous environments}
The largest rate uncertainties for stellar-origin DCO mergers are those
associated with the gaseous disks that power AGN. Circumbinary gas can aid in
the merger of BH-BH binaries in addition to the deep potential wells associated
with nuclear star clusters that reduce the escape of hierarchical GW sources
\citep{McKernan2018, Tagawa2020, Tagawa2021}. The existence of migration traps
is another potential avenue that could lead to significant enhancements in the
rate of BH-hosting DCO mergers \citep[e.g.][]{Bellovary2016}. However, the
lifetimes and properties of AGN disks remain uncertain as do the rate of
prograde and retrograde orbits which can significantly shorten the merger
timescales \citep[][]{Grobner2020} and potentially drive binaries through
inclinations that could excite Lidov-Kozai cycles which further enhance the
merger rates \citep{Dittmann2024}. All of this leads to rates of BH-BH mergers
in AGN disks being predicted to exceed the DCO rate originating in nuclear star
clusters without disks under the assumption that the aforementioned binary
hardening mechanisms are efficient in all cases \citep{Ford2022}. That said, the
evolution of stars in the presence of the gas fueling AGN disks remains a
significant uncertainty that should be explored further \citep{Cantiello2021,
Dittmann2021, Dittmann2024b}. Finally, DCO merger rate uncertainties for
populations originating from circumbinary disks caused by non-accreted material
during Roche overflow is less explored in the literature, but could likely build
off of the work being done for AGN formation channels.

\subsection{Redshift evolution of merger rates}
While Figure~\ref{fig:dco-rates} shows the local DCO merger rates (i.e. at
$z\sim0$), some studies have shown that \emph{redshift-dependent} rates may hold
the key to unraveling the relative contribution of GW sources from different
environments. For example, \cite{Ng2021} showed that BH-BH mergers originating
in isolated binaries, population III binaries (formed from the first stars), and
globular clusters are expected to have merger times that become separable at
redshifts beyond $z\sim7$. When combined with population analyses that
incorporate correlations between measured GW parameters and redshift \citep[e.g.
spin and redshift][]{Biscoveanu2022}, it may be possible to disentangle the
current overlapping predictions from each formation environment. Of course these
distant redshifts are not attainable by current ground-based GW detectors, thus
we must wait until the next crop of ground-based detectors, the so-called XG
class, are realized beyond the 2020s decade \citep{Evans2023, Branchesi2023}.

\subsection{Low frequency gravitational wave source populations}
In contrast to the kHz frequencies that ground-based GW detectors are sensitive
to, space-based GW detectors like LISA will be sensitive to a wide variety of
stellar origin sources containing all combinations of stellar remnants with mHz
orbital frequencies. At lower GW frequencies, the detection horizon of each
source is greatly reduced such that the vast majority of detections are expected
to be made within the Galaxy or local group. In this section, we will focus on
predictions for Galactic populations but note that predictions for
extra-Galactic populations suggest that a small but non-negligible population of
sources should be discoverable by LISA in the local group \citep{Korol2020,
Keim2023} including Andromeda \citep{Korol2018}. This is especially the case for
BH-hosting binaries which could be observed out to Gpc scales depending on the
BH mass. 

\begin{figure}
  \centering
  \includegraphics[width=1\textwidth]{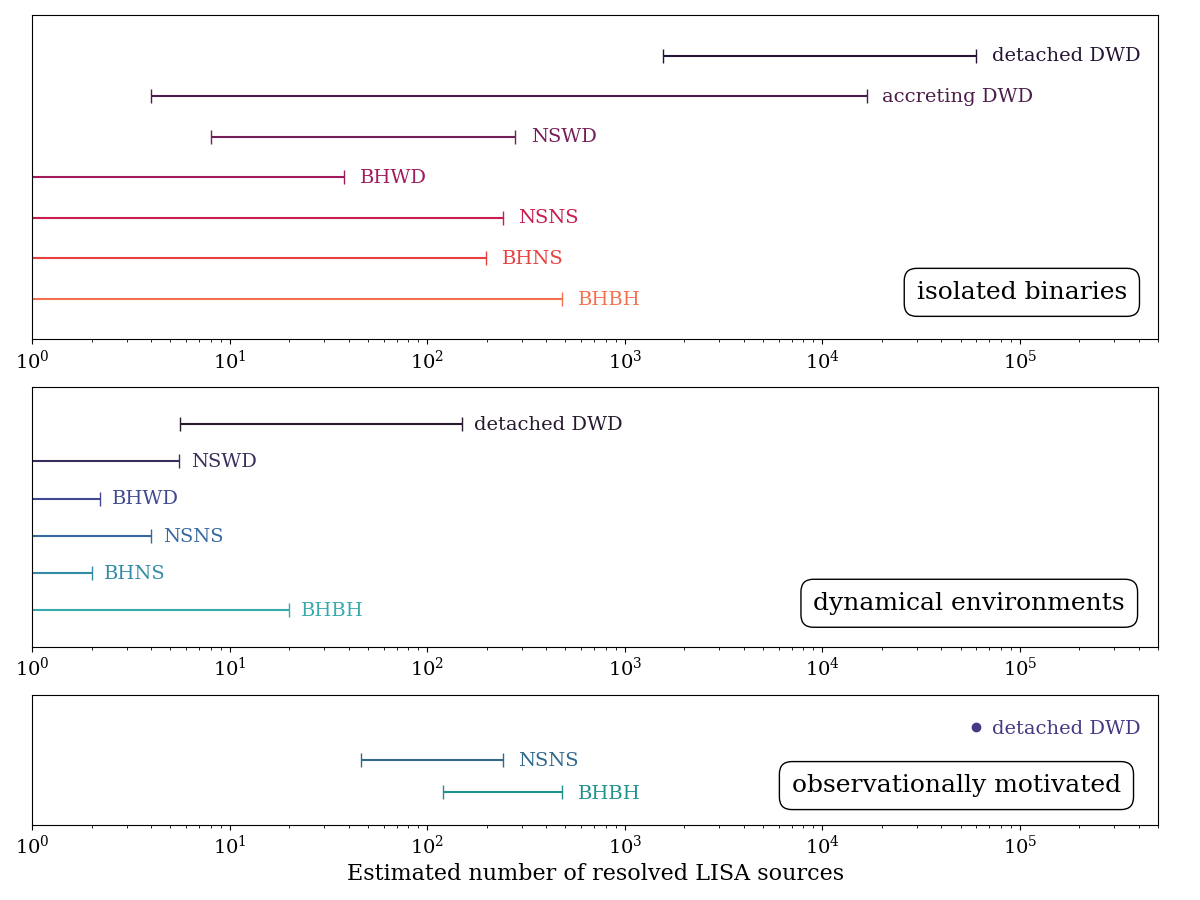}
  
  \caption{Predicted number of sources resolved by LISA for different source classes (colors), environments (top two panels), and prediction strategies (bottom panel). Broadly, WD hosting binaries are the largest resolved source class and isolated binaries are expected to produce the majority of resolved sources. However, we note that in the case of BH-BH sources, only the Galactic globular cluster and nuclear star cluster are considered so the quoted population sizes should be seen as a lower limit.}
  \label{fig:lisa-rates}
\end{figure}

Figure~\ref{fig:lisa-rates} shows the range of predicted source populations
expected to be individually resolved by LISA and originating in isolated
binaries or dynamical formation environments. The lowest panel shows ranges
based on observed rates through other survey strategies.
Table~\ref{tbl:LISA-rates} shows the ranges predicted for a selection of recent
papers using modern population synthesis tools. We refer the reader to each
individual study for a discussion of the assumptions for both the LISA mission
duration and sensitivity as well as the evolutionary assumptions for each
environment. However we note that LEGWORK, a Python-based analytic signal to
noise ratio calculator, has been used in several papers that were published
following its release \citep{legworkapj,legworkjoss}. Below, we discuss the
major uncertainties that affect the predictions of each source class and/or
environment. 

\begin{table}[]
  \TBL{\caption{Predictions for the size of individually
  resolved\footnotemark{a} populations to be discovered by
  LISA}\label{tbl:LISA-rates}}{
    \centering
  \begin{tabular}{lcccccccc}
  \toprule
  
  Reference & Code & Detached WD-WD & Accreting WD & WD-NS & WD-BH & NS-NS &
  NS-BH & BH-BH \\

  \midrule
  \multicolumn{9}{c}{Isolated binaries}\\
  \midrule
  \cite{Nelemans2001b} & SeBa & 5943 &  & 124 & 3 & 39 & 3 & 0 \\
  \cite{Nelemans2004} & SeBa &  & 11000 &  &  &  &  &  \\
  \cite{Ruiter2010} & StarTrack & 6441 & 4859 &  &  &  &  &  \\
  \cite{Belczynski2010} & StarTrack &  &  &  &  & 1.7-4 & 0-0.2 & 0-2.3 \\
  \cite{Yu2010} & BSE & 33670 &  &  &  &  &  &  \\
  \cite{Liu2010} & BSE & 10985 &  &  &  &  &  &  \\
  \cite{Nissanke2012} & SeBa & 1551-6337 & 4-727 &  &  &  &  &  \\
  \cite{Yu2013} & BSE & 3840-19820 &  &  &  &  &  &  \\
  \cite{Liu2014} & BSE &  &  & 132 & 0 & 16 & 3 & 6 \\
  \cite{Yu2015} & BSE &  &  &  & 0-50 &  &  &  \\
  \cite{Kremer2017} & COSMIC &  & 1307-16720 &  &  &  &  &  \\
  \cite{Korol2017} & SeBa & 24482-25754 &  &  &  &  &  &  \\
  \cite{Breivik2018} & COSMIC &  & 77-8295 &  &  &  &  &  \\
  \cite{Lamberts2018} & BSE &  &  &  &  &  &  & 25 \\
  \cite{Lamberts2019} & BSE & 12000 &  &  &  &  &  &  \\
  \cite{Lau2020} & COMPAS &  &  &  &  & 35 &  &  \\
  \cite{Breivik2020a} & COSMIC & 11324 &  & 278 & 0 & 10 & 19 & 72 \\
  \cite{Chen2020} & BSE &  &  & 60-80 &  &  &  &  \\
  \cite{Shao2021} & BSE &  &  &  & 0-28 &  & 2-14 & 12-137 \\
  \cite{Wagg2022} & COMPAS &  &  &  &  & 3-35 & 2-198 & 6-154 \\
  \cite{Liu2022} & BSE/STARS\footnotemark{b} & 75 & 500 &  &  &  &  & \\
  \cite{Biscoveanu2023} & COSMIC &  & 67 - 4045 &  &  &  &  &  \\
  \cite{Li2023} & BSE & 12940-47936 &  &  &  &  &  &  \\
  \cite{Toubiana2024} & BSE &  & 5000-10000 &  &  &  &  &  \\
  \cite{Tang2024} & BPASS &  & 670 &  &  &  &  &  \\
  \cite{Ruiz-Rocha2024} & MOBSE &  &  &  &  &  &  & 2-11 \\
  \midrule
  \multicolumn{9}{c}{Dynamically formed binaries}\\
  \midrule
  \cite{Kremer2018} & CMC & 5.6 & & 5.5 & 2.2 & 1.2 & 0 & 7 \\
  \cite{Kremer2019b} & CMC &  &  &  &  &  &  &  0.5-5 \\
  \cite{Wang2021} & COSMIC\footnotemark{c} & 14-150 & & & & 0.2-4 & 0-2 & 0.3-20
  \\
  \midrule
  \midrule
  \multicolumn{9}{c}{Observationally motivated studies}\\
  \midrule
  \cite{Sesana2016} & None & & & & & &  & 120-480\footnotemark{d} \\
  \cite{Andrews2020} & None & & & & & 46-240 &  & \\
  \cite{Korol2022} & None & 60000\footnotemark{e} &  &  &  &  & & \\
  \botrule
  \end{tabular}}{%
  \begin{tablenotes}
  \footnotetext[a]{We note that each study uses their own signal to noise ratio thresholds and LISA sensitivity curves which have changed over time; we thus encourage the reader to investigate each resource individually}
  \footnotetext[b]{Helium stars with WD companions are considered; the binary
  evolution up to the formation of the AM CVn is computed with BSE while the
  further evolution is computed with an updated version of the STARS code
  \citep{Yungelson2008}.} 
  \footnotetext[c]{The binary evolution was calculated using COSMIC while the dynamical evolution was calculated using custom scripts} 
  \footnotetext[d]{These rates are quoted based on the population rates derived
  in \cite{150914}} \footnotetext[e]{These rates are based on the Type Ia
  population rates derived in \cite{Maoz2018} and references therein}
  \end{tablenotes}}
  \end{table}

\subsubsection{Isolated binaries}
The predicted rate of WD hosting binaries is much higher owing to the impact of
stellar initial mass functions that are heavily weighted toward the lower-mass
WD progenitors \citep[e.g.][]{Kroupa2001}. This is evident in both
Figure~\ref{fig:lisa-rates} and Table~\ref{tbl:LISA-rates}. The major
uncertainties that influence the lower formation rates of NS and BH hosting
binaries detectable by LISA are similar to those discussed above in
Section~\ref{subsubsec:isolated-mergers}. In particular, the strength of natal
kicks at the formation of NS and BH components can lead to a significant
fraction of unbound binaries \cite{Belczynski2010, Wagg2022, Korol2024}. 

In the case of WD-WD binaries, the main uncertainty in the number of predicted
resolved sources lies with the outcomes of Roche-lobe-overflow interactions.
Different assumptions for the stability of mass transfer, which determines
whether Roche-lobe-overflow remains stable or becomes dynamically unstable and
initiates a common envelope, can lead to predicted rate differences within a
factor of $5$ \citep{Thiele2023, Li2023} without changing any other assumptions.
Within a single assumption for mass transfer stability, the choices made for the
envelope ejection efficiency and envelope binding energy can lead to prediction
differences that are orders of magnitude in scale for both detached WD-WD
binaries \citep[e.g.][]{Thiele2023, Korol2024} and accreting WD-WD binaries
\citep[e.g.][]{Kremer2017, Breivik2018,Biscoveanu2023}. In the case of accreting
WD-WD binaries, the stability of mass transfer between WDs adds a further
uncertainty which widens the predicted rates. The use of different star
formation histories for the Galaxy can also lead to wide varieties in the rate
of resolved systems \citep{Yu2013}.

Table~\ref{tbl:LISA-rates} reports predictions for stripped helium stars with WD
companions for \cite{Liu2022} while within the accreting WD classification
helium stars contribute a smaller subpopulation (roughly $5-10\%$) in both
\cite{Nelemans2004} and \cite{Nissanke2012}. We also note that detached stripped
helium star binaries may be detectable by LISA. Indeed, \cite{Gotberg2020} find
that roughly $1-100$ WDs and $0-4$ NSs with stripped-helium-star companions may
be individually resolved by LISA. In a similar context, \cite{Liu2022} find that
up to $75$ detached WDs with stripped-helium-star-companions may be individually
resolved by LISA prior to beginning accretion.

\subsubsection{Dynamically formed binaries}
To date, the only predictions for LISA sources in dynamical environments are
those for the Galactic globular cluster population and the Galactic nuclear star
cluster. Similar to the isolated binaries, dynamically formed binaries hosting
NS and/or BH components are subject to the uncertainties in natal kick strengths
at the compact object formation. If natal kick strengths are large enough that
newly formed COs (and their potential binary hosts) experience velocity
increases of tens of km/s, then the rate of low frequency GW sources can
significantly decrease due to a depletion of COs that can interact within the
Galactic globular clusters. This is less of an issue for the large potential
well that the Galactic nuclear star cluster possesses however. Furthermore, as
with the merger rate predictions for ground-based detectors, the formation rate
and properties of the globular clusters and nuclear star cluster in the Galaxy
add significant uncertainties to the predicted rates of resolved LISA sources
\citep{Kremer2019b,Wang2021}. For WD hosting binaries, the secular evolution of
globular clusters which drives the formation of BH-BH binaries is much less
effective. This leads to a significant reduction in the expected population of
LISA-observable WD-WD binaries that originate in clusters relative to isolated
WD-WD binaries. 

\subsubsection{Observationally motivated predictions}
In recent years, population synthesis calculations have returned to earlier
methods based on the new release of observed datasets. Using the first BH-BH
merger rate estimates made in \cite{150914}, \cite{Sesana2016} showed, under the
assumption that all BH-BH binaries are circular, that LISA should discover a
significant population of BH-BH binaries with mHz GW frequencies. Since 2016,
the BH-BH merger rate has been reduced such that the predictions of
\cite{Sesana2016} should also be reduced from 120-480 to 40-120 resolved BH-BH
binaries discovered by LISA. Based on the Galactic population of NS-NS binaries
discovered through radio surveys and the NS-NS merger rates estimated by
ground-based surveys, \cite{Andrews2020} inferred upper and lower limits for the
LISA resolvable population. They note that the discrepancy between the two
survey strategies likely originates from doppler smearing that makes radio
sources in binaries with orbital frequencies in the LISA band difficult to
discover \citep{Pol2022}. Finally, the WD-WD predictions of \cite{Korol2022}
depend on the assumption that Type Ia supernovae are predominantly formed
through the double degenerate channel and thus directly connected to the WD-WD
population that LISA will discover. They find a rate higher than any theoretical
predictions to date indicating either that the Type Ia mechanism is not wholly
dominated by double degenerate (WD-WD) channels or that LISA will confirm this
to be the case after launch. 

In addition to the studies described in Figure~\ref{fig:lisa-rates} and
Table~\ref{tbl:LISA-rates}, we also note that the population of nearby
cataclysmic variable binaries containing WDs accreting from a hydrogen rich
companion are expected to be observable as a GW foreground \citep{Scaringi2023}.
They build on the volume-limited sample of \cite{Pala2020} and capitalize on the
high space densities of cataclysmic variables in the solar neighborhood combined
with the well known pileup of these binaries at roughly $80\,\rm{min}$ periods
when they experience the so-called `period bounce' \citep{Gansicke2009}. The
period bounce feature is expected to show up in LISA's unresolved Galactic
foreground near $0.1\,\rm{mHz}$.

\section{Properties of GW populations across astrophysical environments}

In this section we describe the properties of simulated GW populations resulting
from population synthesis calculations across the wide variety of astrophysical
environments in which GW sources can form. Rather than subdividing into each
astrophysical environment, we instead discuss the features imprinted in the GW
observables by each environment. 

\subsection{Mass and mass ratio distributions}
Perhaps the largest division between stellar remnants originating in isolated
binary systems and those that are dynamically formed is the ability for
components that form through previous mergers. This is most obviously seen
through BH-hosting binaries with BHs that have masses in the so-called pair
instability supernova mass gap \citep[e.g.][]{Woosley2017}. While the exact
borders of the pair instability mass gap are uncertain, the gap is an
unavoidable feature for BH binaries that originate from stars
\citep{Farmer2020}. Thus, any GW sources with masses in the gap can be assumed
to be formed in dynamical environments
\citep[e.g.][]{Rodriguez2019,Mapelli2021,Tagawa2021}.

Beyond the pair instability mass gap, the mass ratio distribution of BH-BH
binaries has a characteristic shape for systems born in different formation
environments. In the case of triple systems, the mass ratio distribution is
expected to be flat and extend to lower values than mergers formed in isolation
or in globular clusters \citep{Martinez2020}. The most extreme mass ratio
distributions are expected to originate from migration traps in AGN disks
\citep{Secunda2020}, however we note that recent discoveries of BHs with
luminous stellar companions in the Galaxy also have extreme mass ratios of
$q\sim1/30-1/10$ \citep{BH1, Chakrabarti2023, BH2, BH3}.

\subsection{Spins}
The spins of BHs formed in different astrophysical environments are delineated
by whether the merger components are born and evolve together or whether they
are dynamically assembled. In the case of dynamically assembled mergers, the
spins, regardless of their magnitude should be oriented completely randomly
\citep{Rodriguez2016b,Antonini2019}. This is less obviously the case for
isolated binaries which may have spins aligned with the orbital angular momentum
vector as expected from the chemically homogeneous channel
\citep{Marchant2016,Mandel2016} or could have spins that are not aligned due to
strong natal kicks \citep{Baibhav2023}. Triple systems are also more likely to
have aligned spins, thus leading to a potential breaking of the degeneracy
between clusters, isolated binaries, and triple systems \citep{Martinez2020}.
Finally, in the case of gas-driven mergers, the spin distributions can vary
based on whether the predominant channels for merging BH-BH binaries originate
in prograde or retrograde orbits or in migration traps \citep{McKernan2022}.

\subsection{Eccentricity}
The vast majority of GW sources are expected to have immeasurably low
eccentricities by the time they reach frequencies that ground-based detectors
are sensitive to. In a few cases, high eccentricities can be driven through
extended dynamical interactions like binary-binary encounters \citep{Zevin2019},
repeated mergers in highly concentrated globular clusters \citep{Rodriguez2019},
or eccentric Kozai-Lidov cycles \citep{Wang2021}. These systems may indeed merge
with eccentricities that could be detectable by ground-based GW detectors and
their absence may play a role in our understanding of the relative contribution
of dynamical formation channels \citep{Zevin2021}. However, we note that with
present analysis techniques, it is extremely difficult to disentangle the
effects of eccentricity and spin precession in GW waveforms
\citep{Romero-Shaw2023, Fumagalli2023, Fumagalli2024}. 

At lower frequencies, however, the residual eccentricity imprinted from each
channel may be observable by LISA \citep{Nishizawa2016, Breivik2016}. Indeed,
the degree of dynamical activity is expected to be imprinted with GW sources
residing in stellar clusters having larger eccentricities than GW sources that
have been ejected from the clusters \citep{Kremer2019b}. Even smaller
eccentricities arising from natal kicks at the formation of BHs and NSs in
isolation \citep{Breivik2016, Andrews2020, Korol2024} or circumbinary disk
interactions \citep{Romero-Shaw2024} are expected to be measurable by LISA. For
WD-WD binaries, however, all predictions from isolated binaries result in
perfectly circular systems. Thus, eccentricity will also help to delineate the
dynamically formed WD-WD binaries as well \citep{Willems2007}.

\section{Future directions and Conclusions}
At the time of writing, the LIGO detectors are operating in their fourth
observing run with over one hundred merger candidates reported so far. LISA has
passed mission adoption in early 2024 and XG ground-based detector development
is underway. In many ways, population synthesis studies are expected to continue
in the same fashion as they have since their inception, with population rates
and characteristics reported as new formation channels or interaction physics
models are refined. However, it has become clear that understanding the vast
array of possible formation scenarios for GW sources is a near impossible feat
without having anchors that can be used as foot holds to understand the relative
contribution of each channel. Indeed, if we are missing yet undiscovered
formation channels, our inference of multi-origin GW source populations
\citep[e.g.][]{Wong2021, Zevin2021a} has been shown to be biased
\citep{Cheng2023}.

Some potential pathways for avoiding this bias include better understanding of
all possible uncertainties within a given GW source environment
\citep[e.g.][]{Wong2023} or searching for features that remain constant under
wide parameter assumption variations \citep[][]{vanSon2022, vanSon2023}. In
addition to these efforts, the application of new data that is complimentary to
GW source populations and captures different phases of evolution across each
environment may play a central role in future investigations. Indeed, the
discoveries of the first three BH binaries with luminous companions based on
Gaia data show a wide range of mass and composition \citep{BH1, Chakrabarti2023,
BH2, BH3}. It is thus difficult to understate the value that future detections
from Gaia's upcoming data releases may hold
\citep[e.g.][]{Breivik2017,Chawla2022,Janssens2022}. The future of astrometric
detection of Galactic binary systems is similarly bright for WD and NS binaries
\citep{Shahaf2024, Yamaguchi2024, El-Badry2024}. While these sources are not
necessarily expected to be the direct progenitors of all GW sources, they
provide critical datasets that can be used to constrain the outcomes of binary
interactions which play a central role in the production of nearly all potential
GW sources. 

Finally, as SDSS-V and the Vera C. Rubin Legacy Survey of Space and Time release
data in the coming years, discoveries through radial velocity and photometric
variations are expected to be similarly abundant \citep{Weller2023, Chawla2023}.
The combined analysis of binary and higher-order multiple systems in the Galaxy
and local group with GW sources is sure to provide cross-cutting discoveries
that could aid in breaking the degeneracies currently restricting GW population
analyses. These analyses will be bolstered by continued development support and
use of open-source population synthesis tools that span all astrophysical
environments.

\begin{ack}[Acknowledgments]
  KKB is grateful for helpful conversations on early population synthesis
 predictions and their implementation with Lev Yungelson, Alexander Tutukov, and
 Alexey Bogomazov. 
\end{ack}


\bibliographystyle{Harvard}
\bibliography{reference}

\end{document}